\documentstyle[aps,preprint,eqsecnum,floats,epsf]{revtex}
\tighten
\draft
\begin{document}

\preprint{}
\title{Generalized Ladder Operators for Shape-invariant Potentials}
\author{Elso Drigo Filho\thanks{Electronic address:
         {\tt elso@df.ibilce.unesp.br}} and
M.~A. C\^andido Ribeiro\thanks{Electronic address:
         {\tt macr@df.ibilce.unesp.br}},}
\address{Departamento de F\'{\i}sica - Instituto de Bioci\^encias,
         Letras e Ci\^encias Exatas - 
         UNESP \\ S\~ao Jos\'e do Rio Preto, SP 15054-000 - Brazil}
\date{\today}
\maketitle

\begin{abstract}

  A general form for ladder operators is used to construct a method to
  solve  bound-state Schr\"odinger  equations. The  characteristics of
  supersymmetry and shape invariance of the system are the start point
  of the approach. To show the  elegance and the utility of the method
  we  use it  to  obtain  energy spectra  and  eigenfunctions for  the
  one-dimensional harmonic oscillator and Morse potentials and for the
  radial harmonic oscillator and Coulomb potentials.
\end{abstract}


\section{Introduction}

Algebraic methods  applied to quantum mechanical problems  have a long
history  \cite{ref1}. The  development of  the  supersymmetric quantum
mechanics   \cite{ref2',ref2}  has   contributed   to  enlarge   these
approaches.  Examples of  this contribution  is the  concept  of shape
invariance introduced  by Gendenshtein  \cite{ref3}, and the  study of
coherent  states for one-dimensional  potentials using  a superalgebra
formalism \cite{ref4,ref5,ref6}.

Recently, Balantekin has shown that shape invariance has an underlying
algebraic structure \cite{ref7}. That  approach was applied to studied
barrier penetration \cite{ref8}  and exactly solvable coupled-channels
problems \cite{ref9}.

In  this article  we show  that the  algebraic structure  presented by
Balantekin can  be applied as a  generalized method to  solve one- and
three-dimensional Schr\"odinger equations exactly. When one introduces
shape invariance  added to supersymmetric formalism it  is possible to
obtain   a   general   approach   to  construct   generalized   ladder
operators.   The  one-dimensional   harmonic   oscillator  and   Morse
potential,  the  radial equation  for  the three-dimensional  harmonic
oscillator  and Coulomb  potential are  analyzed using  this algebraic
approach.  Different  forms of  raising  and  lowering operators,  not
defined  in   the  shape   invariance  context,  have   been  proposed
\cite{ref10,ref11,ref12} for these potentials.

The usual approach to solve Schr\"odinger equations using superalgebra
formalism   is   by   constructing   a   hierarchy   of   Hamiltonians
\cite{ref12'}. For  the method showed in this  paper that construction
is not necessary. The $n$-th eigenfunction for a given shape invariant
potential can be  deduced from the ground state  wave function and the
appropriated ladder operators.

In  the next  section  we  introduce the  formalism  to construct  the
generalized ladder  operators. In the section  III the one-dimensional
potentials are  solved using this approach, as  examples, the harmonic
oscillator  and the  Morse  potential are  explicitly  solved. In  the
section  IV  the radial  Schr\"odinger  equation  is  studied and  the
spectrum  and   eigenfunctions  for  the   three-dimensional  harmonic
oscillator and Coulomb potential are obtained.

\section{The Generalized Ladder Operators}

From a bound-state Hamiltonian as ($\hbar$ and $2m=1$)

\begin{equation}
H=-{d^2\over dx^2}+V(x)\;
\label{eqH}
\end{equation}
with

\begin{equation}
H \Psi_n(x) = E_n \Psi_n (x)\;,
\label{eqHP}
\end{equation}
the supersymmetric quantum mechanics can be constructed if $H$ can be
factorized in the bosonic operators:

\begin{equation}
A^\pm(x;a)=\mp{d\over dx}+W(x;a)\;,
\label{eqA+-}
\end{equation}
where  $W(x;a)$  is well  known  as  the  superpotential, which  is  a
function of the  position variable and a set  of parameters, $a$, that
represent  space-independent  properties  of  the  original  potential
$V(x)$. If so, we obtain the first supersymmetric partner Hamiltonian

\begin{equation}
H_+= H-E_0=A^+A^-\;,
\label{eqH+}
\end{equation}
where $E_0$ is the ground state energy eigenvalue. The second partner
Hamiltonian is

\begin{equation}
H_-\equiv A^-A^+\;.
\label{eqH-}
\end{equation}
The Hamiltonians $H_+$  and $H_-$ have the same  energy spectra except
the ground state  of $H_+$, for which there  is no corresponding state
in the  spectra of $H_-$. We  can see that,  from Eq.~(\ref{eqH+}) the
superpotential satisfies the Ricatti equation:

\begin{equation}
W(x;a)-W^\prime(x;a)=V(x)-E_0=V_+(x)\;.
\label{eqRic}
\end{equation}

It  was  introduced in  1983  by  Gendenshtein  \cite{ref3} the  shape
invariance condition

\begin{equation}
R(a_1)=V_-(x;a_0)-V_+(x;a_1)\;,
\label{eqR}
\end{equation}
where $R(a_1)$ is independent of any dynamical variable and $a_1$ is a
function of $a_0$. Potentials which satisfy this condition are exactly
solvable.  However,   shape  invariance   is  not  the   most  general
integrability  condition, once there  are exactly  solvable potentials
that seen do not show  the shape invariance property \cite{ref14}.  In
this work  we consider shape invariance involving  translations of the
parameters $a$:

\begin{equation}
a_1=a_0+\eta \;,
\label{eqt}
\end{equation}
where  $\eta$ is  the translation  step.  The other  kind of  relation
between the  parameters so far  studied, scaling \cite{ref15},  is not
broached here because only for the case of translations the potentials
are explicitly known  and can be written in a closed  form in terms of
simple functions, whereas  for the case of scaling  the potentials can
only be  written as a  Taylor series \cite{ref2}.  In this way,  it is
possible to define translation operators $T(a_0)$ as

\begin{mathletters}
\label{eqT}
\begin{eqnarray}
& & T(a_0) = \exp \left( \eta {\partial \over \partial a_0}\right)\\ &
& T^{-1}(a_0)= T^{\dagger}(a_0) = \exp \left( -\eta {\partial \over
\partial a_0}\right)\,.
\end{eqnarray}
\end{mathletters}
These operators only act on objects defined in the parameters space.

In  order  to  introduce  the generalized  creation  and  annihilation
operators  we compose the  translation operators  $T$ and  the bosonic
operators $A$ as

\begin{mathletters}
\label{eqB}
\begin{eqnarray}
& & B_+(a_0) = A^+(a_0)T(a_0)\\ &
& B_-(a_0) = T^{\dagger}(a_0)A^-(a_0)\,.
\end{eqnarray}
\end{mathletters}
These  operators   $B$  present  the   necessary  algebraic  structure
\cite{ref7} to  identify them  as ladder operators.  As such  they are
generalizations of the harmonic oscillator ladder operators.

The  method  to  solve  Schr\"odinger  equation  with  shape-invariant
potentials is similar to a factorization method applied to the case of
a  harmonic oscillator  potential. Thus,  the ground  state,  $ \Psi_0
(x;a_0)$, must obey

\begin{equation}
B_-(a_0)\Psi_0 (x;a_0) = A^-(a_0)  \Psi_0 (x;a_0) =0\;,
\label{eqgs}
\end{equation}
or

\begin{equation}
 \Psi_0 (x;a_0) \propto \exp \left(-\int^x W(\bar x;a_0)d\bar x
\right)\;.
\label{eqgsi}
\end{equation}
Eq.~(\ref{eqgsi})  is a  relation  between the  ground  state and  the
superpotential   commonly  found   in   the  supersymmetric   approach
\cite{ref2,ref12'}. The  excited states  are obtained by  the repeated
action of the creation operator on the ground state

\begin{equation}
\Psi_n (x;a_0) = [B_+(a_0)]^n  \Psi_0 (x;a_0) \;.
\label{eqes}
\end{equation}
At this point  we want to stress that, since the  operators $T$ do not
commute with  any $a$-dependent  object and the  operators $A$  do not
commute with any  $x$-dependent object then, the operators  $B$ act on
objects defined  in the dynamical  variable space only by  the bosonic
operators,  and on  objects defined  in  the parameters  space by  the
translation  operators.  The  normalization  constant  that  transform
Eq.~(\ref{eqgsi}) into a relation  of equality, in general, depends on
the parameters $a$. Thus, this constant is also affected by the action
of the operators $B$ on the eigenfunctions. The normalization constant
is not explicitly computed in this work, it can be included during the
process of the use of the operators $B$ \cite{ref6} or determined from
the normalization condition of the eigenfunctions.

In  order to  determine the  energy  eigenvalues, it  is important  to
understand the rule played by the remainder $R(a_1)$, Eq.~(\ref{eqR}),
in   this  context.   From  the   translation  operators   defined  in
Eqs.~(\ref{eqT}), we can write

\begin{equation}
R(a_n)=T(a_0)R(a_{n-1})T^{\dagger}(a_0)\;,                        
\label{eqRn}
\end{equation}
where

\begin{equation}
a_n=a_0+n\eta \;,
\label{eqan}
\end{equation}
is  a generalization  of the  relation given  by  Eq.~(\ref{eqt}). The
superalgebra permits  to obtain a  simple relation between  the energy
spectrum and the remainders $R(a_n)$. To see how it is possible we use
Eq.~(\ref{eqRn}) to write down

\begin{equation}
R(a_n)B_+(a_0)=B_+(a_0)R(a_{n-1})\;.                        
\label{eqRnB+}
\end{equation}
From  Eqs.~(\ref{eqRn})  and  (\ref{eqRnB+})  we  have  the  following
commutation relation

\begin{equation}
[H_+,(B_+)^n]=\left( \sum_{k=1}^{n}R(a_k)\right)(B_+)^n\;.
\label{eqH+B+}
\end{equation}
Now we apply this equation on $\Psi_0 (x;a_0)$ to have

\begin{equation}
H_+\left\{ [B_+(a_0)]^n \Psi_0 (x;a_0)\right\}=\left(
\sum_{k=1}^{n}R(a_k)\right)\left\{ [B_+(a_0)]^n \Psi_0 (x;a_0)\right\}\;,
\label{eqH+B+Psi}
\end{equation}
since  $\Psi_0 (x;a_0)$  is  the ground  state  of $H_+$  with a  null
eigenvalue. Thus, $[B_+(a_0)]^n \Psi_0 (x;a_0)$ is an eigenfunction of
$H_+$ with  eigenvalue $\epsilon_n=\sum_{k=1}^{n}R(a_k)$. Therefore we
have for the energy spectrum

\begin{equation}
E_n=E_0+\epsilon_n=E_0+\sum_{k=1}^{n}R(a_k)\;,                        
\label{eqEn}
\end{equation}
where $E_0$  can be  obtained directly from  the factorization  of the
original Hamiltonian, Eq.~(\ref{eqH+}) or Eq.~(\ref{eqRic}).

We would like to close this section by emphasizing that this algebraic
approach  is self-consistent and  it permits  to determine  the energy
eigenvalues and eigenfunctions of a bound-state Schr\"odinger equation
from supersymmetric and shape  invariance properties of the system. In
this  way this  approach  is a  method  for exact  resolution of  that
equation.

\section{One-dimensional Applications}

To illustrate the  method discussed in the preceding  section we first
analyse one-dimensional shape-invariant potentials.

\subsection{Harmonic Oscillator}

This is the  simplest example of a shape-invariant  potential. In this
sense, it is interesting in  order to become the approach clearer. The
original Hamiltonian is

\begin{equation}
H=-{d^2\over dx^2}+x^2\;
\label{eqHho}
\end{equation}
and the factorization is 

\begin{equation}
H_+=A^+A^-=\left(-{d\over dx}+ax\right)\left({d\over dx}+ax\right)
=-{d^2\over dx^2}+a^2x^2-a\;,
\label{eqH+ho}
\end{equation}
thus the superpotential is simply given by

\begin{equation}
W(x;a)=ax\;.
\label{eqWho}
\end{equation}
The supersymmetric partner Hamiltonian is

\begin{equation}
H_-=A^-A^+=-{d^2\over dx^2}+a^2x^2+a\;,
\label{eqH-ho}
\end{equation}
and the shape invariance condition, Eq.~(\ref{eqR}), establishes that

\begin{equation}
R(a_1)=a_0^2x^2+a_0-a_1^2x^2+a_1\;.
\label{eqRho}
\end{equation}
Since $R(a_1)$  must not depend  on the $x$-coordinate  then $a_0=a_1$
and  a  comparison of  $H_+$,  Eq.~(\ref{eqH+ho}),  with the  original
Hamiltonian $H$,  Eq.~(\ref{eqHho}), results  that, in this  case, all
parameters are equal to 1. So we get $a_n=a_0=1$, $\eta=0$, $R(a_k)=2$
and for the ground state energy, that is given by the constant term in
Eq.~(\ref{eqH+ho}), $E_0=1$.

The  harmonic oscillator  is the  simplest case  of  a shape-invariant
potential  because the  parameter  step is  null  ($\eta=0$) and  from
Eqs.~(\ref{eqT}) and (\ref{eqB})  the generalized ladder operators are
simply  the same  as the  bosonic ones  ($B_{\pm}=A^{\pm}$).  Thus, we
recover the  standard approach of the ladder  operators method applied
to the  harmonic oscillator. In  this way, from  Eqs.~(\ref{eqgs}) and
(\ref{eqEn}) we have

\begin{equation}
B_-(a_0) \Psi_0 (x;a_0) = A^-(a_0)  \Psi_0 (x;a_0) =0\;\;\;\;
\Longrightarrow \;\;\;\;  \Psi_0 (x;a_0) \propto {\mbox
e}^{-x^2/2}\;,
\label{eqgsho}
\end{equation}

\begin{equation}
E_n=E_0+\sum_{k=1}^{n}R(a_k)=1+\sum_{k=1}^{n}2=2n+1\;,                       
\label{eqEnho}
\end{equation}
as it should be.

\subsection{Morse Potential}

Algebraic approaches,  but not developed  within the context  of shape
invariance,  for  one-dimensional  Morse  potential  can  be  find  in
Refs.~\cite{ref10,ref12,ref17}. In Ref.~\cite{ref7} it is explored the
shape invariance  property to determine  the SU(1,1) structure  of the
algebra associated to that potential.

Here we start with the  original bound-state Hamiltonian for the Morse
potential,

\begin{equation}
H=-{d^2\over dx^2}+\lambda^2\left( 1-{\mbox e}^{-x}\right)^2\;.
\label{eqHMp}
\end{equation}
Factorization of this Hamiltonian gives

\begin{eqnarray}
H_+=H-E_0&=&\left(-{d\over dx}+\lambda\left( 1-{\mbox
e}^{-x}\right)-a\right)\left({d\over dx}+\lambda\left( 1-{\mbox
e}^{-x}\right)-a\right) \nonumber \\
&=&-{d^2\over dx^2}+\lambda^2\left( 1-{\mbox e}^{-x}\right)^2
-2a\lambda +a^2+ \lambda\left( 2a-1\right){\mbox e}^{-x}\;,
\label{eqH+Mp}
\end{eqnarray}
where the superpotential is

\begin{equation}
W(x;a)=\lambda\left( 1-{\mbox e}^{-x}\right)-a\;.
\label{eqWMp}
\end{equation}
The supersymmetric partner of $H_+$ is

\begin{equation}
H_-=A^-A^+=-{d^2\over dx^2}+\lambda^2\left( 1-{\mbox e}^{-x}\right)^2
-2a\lambda +a^2+ \lambda\left( 2a+1\right){\mbox e}^{-x}\;.
\label{eqH-Mp}
\end{equation}
The shape invariance condition, Eq.~(\ref{eqR}), for this potential
can be written as

\begin{equation}
R(a_1)=\left( 2a_0-2a_1+2\right)\lambda {\mbox e}^{-x}-2\left(
a_0-a_1\right)\lambda -a_1^2+a_0^2\;.
\label{eqRMp}
\end{equation}
In  order to  have $R(a_1)$  independent of  the variable  $x$,  it is
necessary  that $a_1=a_0+1$  then, from  Eq.~(\ref{eqt}),  $\eta=1$. A
comparison between Eq.~(\ref{eqH+Mp}) and Eq.~(\ref{eqHMp}) shows that
$a_0=1/2$ and $E_0=\lambda -1/4$. The shape invariance condition gives
to $R(a_k)$ the following form:

\begin{equation}
R(a_k)=-2\left( a_{k-1}-a_k\right)\lambda -a_k^2+a_{k-1}^2
=2\lambda-2k\;.
\label{eqRkMp}
\end{equation}
For this case the energy spectrum, given by Eq.~(\ref{eqEn}), is

\begin{equation}
E_n=E_0+\sum_{k=1}^{n}\left( 2\lambda-2k\right)
=2\lambda\left(n+{1\over 2} \right)-\left( n+{1\over 2}\right)^2 \;,
\label{eqEnMp}
\end{equation}
that is the exact result \cite{ref11,ref12}.

From Eqs.~(\ref{eqB}), the generalized ladder operators can be written
as

\begin{mathletters}
\begin{eqnarray}
& & B_+(a_0) = A^+(a_0)T(a_0)= 
\left[ -{d\over dx}+\lambda\left(
1-{\mbox e}^{-x}\right)-a_0\right]\exp \left(
{\partial \over \partial a_0}\right)
\label{eqBMpa}\\ 
& & B_-(a_0) = T^{\dagger}(a_0)A^-(a_0)= 
\exp \left( -{\partial \over \partial a_0}\right)
\left[ {d\over dx}+\lambda\left( 1-{\mbox 
e}^{-x}\right)-a_0\right]\,.
\label{eqBMpb}
\end{eqnarray}
\end{mathletters}
Substitution  of Eq.~(\ref{eqWMp})  into Eq.~(\ref{eqgsi}),  gives for
the ground state eigenfunction the following expression

\begin{equation}
\Psi_0 (x;a_0) \propto {\mbox e}^{a_0x}\;{\exp}\left[{-\lambda \left(
x+{\mbox e}^{-x}\right)}\right]\;.
\label{eqgsMp}
\end{equation}
The  excited states  are constructed  by  the repeated  action of  the
creation  operator  $B_+$,  Eq.~(\ref{eqBMpa}),  on the  ground  state
Eq.~(\ref{eqgsMp}). For example, for the first excited state we have

\begin{equation}
\Psi_1 (x;a_0)\propto B_+(a_0)\Psi_0 (x;a_0)=  
{\exp}\left[ \left( -\lambda + {a_0+1}\right)x\right]
\;{\exp}\left[{-\lambda {\mbox e}^{-x}}\right]\;
\left\{ 2\lambda -2\lambda {\mbox e}^{-x}
-2a_0 -1\right\}.
\label{eqfsMp}
\end{equation}
These same results for the Morse potential eigenfunctions \cite{ref11}
can  be  obtained  by  other  exact  methods,  just  remembering  that
$a_0=1/2$.

\section{Three-Dimensional Potentials}

The algebraic approach  developed in the section II  is not restricted
to one-dimensional potentials. As a matter of fact, it can immediately
be extended to  radial equations for central force  potentials in more
dimensions. In particular, here we  present the results for the radial
harmonic oscillator and the Coulomb potentials in three dimensions.

\subsection{Radial Harmonic Oscillator}

The original Hamiltonian for a radial harmonic oscillator is

\begin{equation}
H=-{d^2\over dr^2}+r^2+{l(l+1)\over r^2}\;.
\label{eqHrho}
\end{equation}
Following  the same  development  used in  one-dimensional cases,  the
Hamiltonian given  by Eq.~(\ref{eqHrho})  can be factorized,  for each
value of $l$, as

\begin{eqnarray}
H_+=H-E_0^{(l)}&=&\left(-{d\over dr}+r-{l+1\over r}\right)
\left({d\over dr}+r-{l+1\over r}\right) \nonumber \\
&=&-{d^2\over dr^2}+r^2+{l(l+1)\over r^2}-2l-3\;,
\label{eqH+rho}
\end{eqnarray}
where the superpotential is

\begin{equation}
W(r;l)=r-{l+1\over r}\;.
\label{eqWrho}
\end{equation}
The supersymmetric partner of $H_+$, Eq.~(\ref{eqH+rho}), is

\begin{equation}
H_-=A^-A^+=-{d^2\over dr^2}+r^2+{(l+1)(l+2)\over r^2}-2l-1\;.
\label{eqH-rho}
\end{equation}
For this case, the shape invariance condition, Eq.~(\ref{eqR}), can be
written as

\begin{equation}
R(l_1)=r^2+{(l_0+1)(l_0+2)\over r^2}-2l_0-1-r^2-{l_1(l_1+1)\over
r^2}+2l_1+3\;.
\label{eqRrho}
\end{equation}
Note that here, the parameter  $a$ gives place to the angular momentum
number   $l$.   Imposed   that    the   right-hand   side   terms   of
Eq.~(\ref{eqRrho})  must not  depend on  the variable  $r$,  we obtain
$l_1=l_0+1$.    Then,   $\eta   =1$    and   $R(l_1)=4$,    and   from
Eq.~(\ref{eqH+rho})   $E_0^{(l)}=2l+3$.  As  in   the  one-dimensional
harmonic oscillator case, the function $R(l_k)$ is constant. Thus, the
energy spectrum for the 3-D harmonic oscillator is

\begin{equation}
E_n^{(l)}=E_0^{(l)}+\sum_{k=1}^{n}R(l_k)=2l+3+4n\;.                       
\label{eqEnrho}
\end{equation}

The  eigenfunctions  can  be  obtained  from  the  generalized  ladder
operators:

\begin{mathletters}
\begin{eqnarray}
& & B_+(l) = A^+(l)T(l)= 
\left[ -{d\over dr}+ r-{l+1\over r}\right]\exp \left(
{\partial \over \partial l}\right)
\label{eqBrhoa}\\ 
& & B_-(l) = T^{\dagger}(l)A^-(l)= 
\exp \left( -{\partial \over \partial l}\right)
\left[ {d\over dr}+ r-{l+1\over r}\right]\,.
\label{eqBrhob}
\end{eqnarray}
\end{mathletters}
The minimum energy  state for a fixed $l$ ($n=0$),  is obtained by the
substitution of Eq.~(\ref{eqBrhob}) into Eq.~(\ref{eqgs}), giving

\begin{equation}
\Psi_0^{(l)} (r) \propto r^{l+1}\;{\mbox e}^{-r^2/2}\;.
\label{eqgsrho}
\end{equation}
The other eigenfunctions  (for $n\geq 1$) can be  obtained by repeated
action of  the creation operator, Eq.~(\ref{eqBrhoa})  on the function
given by Eq.~(\ref{eqgsrho}). As an example, we have for $n=1$

\begin{equation}
\Psi_1^{(l)} (r)\propto B_+(l)\Psi_0^{(l)} (r)=  
\left( 2r^2-2l-3\right)r^{l+1}\;{\mbox e}^{-r^2/2}\;.
\label{eqfsrho}
\end{equation}
These  results  are  obtained  by  other  methods,  for  instance,  in
Ref.~\cite{ref18}.

\subsection{Radial Coulomb Potential}

The radial Hamiltonian for the Coulomb potential is
 
\begin{equation}
H=-{d^2\over dr^2}-{1\over r}+{l(l+1)\over r^2}\;,
\label{eqHrCp}
\end{equation}
which is factorized as

\begin{eqnarray}
H_+=H-E_0^{(l)}&=&\left(-{d\over dr}+{1\over 2(l+1)}-{l+1\over r}\right)
\left({d\over dr}+{1\over 2(l+1)}-{l+1\over r}\right) \nonumber \\
&=&-{d^2\over dr^2}-{1\over r}+{l(l+1)\over r^2}+{1\over 4(l+1)^2}\;,
\label{eqH+rCp}
\end{eqnarray}
where 

\begin{equation}
W(r;l)={1\over 2(l+1)}-{l+1\over r}\;.
\label{eqWrCp}
\end{equation} 
The supersymmetric partner of the Hamiltonian (\ref{eqH+rCp}) is

\begin{equation}
H_-=A^-A^+=-{d^2\over dr^2}-{1\over r}+{(l+1)(l+2)\over r^2}
+{1\over 4(l+1)^2}\;.
\label{eqH-rCp}
\end{equation}
Here  again  the quantum  number  $l$ is  used  as  the parameter  $a$
($a_0=l_0$). Using the shape invariance condition, we obtain

\begin{equation}
R(l_1)=-{1\over r}+{(l_0+1)(l_0+2)\over r^2}+{1\over 4(l_0+1)^2}
+{1\over r}-{l_1(l_1+1)\over r^2}-{1\over 4(l_1+1)^2}\;.
\label{eqRrCp}
\end{equation}
In order to have $R(l_1)$ independent on $r$ we must have $l_1=l_0+1$
and thus, $\eta = 1$ and 

\begin{equation}
R(l_1)={1\over 4(l_0+1)^2}-{1\over 4(l_1+1)^2}\;.
\label{eqR1Cp}
\end{equation}
For the $k$-th parameter, the shape invariance condition gives

\begin{eqnarray}
R(l_k)&=&{1\over 4(l_{k-1}+1)^2}-{1\over 4(l_k+1)^2}\nonumber \\
&=&{1\over 4}\left[ {1\over (l+n)^2}-{1\over (l+n+1)^2}\right]\;.
\label{eqRkCp}
\end{eqnarray}
The energy  eigenvalue $E_0^{(l)}$ is obtained from  the constant term
in Eq.~(\ref{eqH+rCp});

\begin{equation}
E_0^{(l)}={1\over 4(l+1)^2}\;.
\label{eqE0rCp}
\end{equation}
Thus, the energy spectrum for this case is given by

\begin{equation}
E_n^{(l)}=E_0^{(l)}+\sum_{k=1}^{n}{1\over 4}\left[ {1\over
(l+n)^2}-{1\over (l+n+1)^2}\right]=
-{1\over 4(n+l+1)^2}\;.
\label{eqEnrCp}
\end{equation}

The generalized ladder operators for the radial Coulomb potential are

\begin{mathletters}
\begin{eqnarray}
& & B_+(l) = A^+(l)T(l)= 
\left[ -{d\over dr}+ {1\over 2(l+1)}-{l+1\over r}\right]\exp \left(
{\partial \over \partial l}\right)
\label{eqBrCpa}\\ 
& & B_-(l) = T^{\dagger}(l)A^-(l)= 
\exp \left( -{\partial \over \partial l}\right)
\left[ {d\over dr}+{1\over 2(l+1)} -{l+1\over r}\right]\,.
\label{eqBrCpb}
\end{eqnarray}
\end{mathletters}
The lowest energy  level eigenfunction for each $l$  is obtained using
Eqs.~(\ref{eqgs})  and  (\ref{eqBrCpb})  or  by  the  substitution  of
Eq.~(\ref{eqWrCp}) into Eq.~(\ref{eqgsi})

\begin{equation}
\Psi_0^{(l)} (r) \propto r^{l+1}\;{\mbox e}^{-r/2(l+1)}\;.
\label{eqgsrCp}
\end{equation}
The other eigenfunctions (for $n\geq  1$) are calculated by the action
of   the   creation  operator,   Eq.~(\ref{eqBrCpa})   on  the   above
eigenfunction.  As  an  example,  for  $n=1$  we  have  the  following
eigenfunction:

\begin{equation}
\Psi_1^{(l)} (r)\propto B_+(l)\Psi_0^{(l)} (r)=  
\left\{ {1\over 2}\left[ {1\over l+2}+{1\over l+1}\right]-2l-3\right\}
r^{l+1}\;{\mbox e}^{-r/2(l+2)}\;.
\label{eqfsrCp}
\end{equation}
These results  for the radial  Coulomb potential are also  obtained by
other exact methods \cite{ref11}. Usually for this case, the principal
quantum number adopted is $N=n+l+1$.

\section{Concluding Remarks}

A  general  constructive  method  to solve  bound-state  Schr\"odinger
equations for shape invariant  potentials is presented. This algebraic
method is based on a generalization of harmonic oscillator raising and
lowering  operators. The  introduction of  hierarchy  of Hamiltonians,
usual in this kind of  algebraic treatment, is no longer necessary for
the   present  approach.   Solutions   for  one-dimensional   harmonic
oscillator  and  Morse  potentials  and for  three-dimensional  radial
harmonic oscillator and Coulomb potentials are explicitly obtained.

Solutions for other shape invariant potentials can also be obtained by
this approach.  In particular, radial harmonic  oscillator and Coulomb
potential  in  arbitrary   dimensions  \cite{ref19,ref20}  are  simple
extensions of the results appointed here.

An   interesting  discussion   is  concerning   coherent   states  for
one-dimensional   potentials  and  supersymmetric   quantum  mechanics
\cite{ref6}.  In this  way one  has a  elegant and  compact  method to
introduce  ladder  operators and  define  coherent  states from  these
operators.

\section*{ACKNOWLEDGMENTS}

The   authors   thank   Prof.   A.  Baha   Balantekin   for   valuable
discussions. M.A.C.R.\  acknowledges the  support of Funda\c  c\~ao de
Amparo  \`a   Pesquisa  do  Estado  de  S\~ao   Paulo  (Contract  No.\
98/13722-2).



\begin{references}

\bibitem{ref1} Schr\"odinger, E., Proc. Roy. Irish Acad. {\bf A46}, 9
(1940); Infeld, L. and Hull, T. E., Rev. Mod. Phys. {\bf 23},
21 (1951).

\bibitem{ref2'} Haymaker, R. W. and Rau, A. R. P., Am. J. Phys. {\bf
54}, 928 (1986).

\bibitem{ref2} Cooper, F., Khare, A. and Sukhatme, U. P., Phys. Rep.
{\bf 251}, 267 (1995).

\bibitem{ref3} Gendenshtein, L., Piz. Zh. Eksp. Teor. Fiz.
{\bf 38}, 299 (1983) (Engl. trans. JETP Lett. {\bf 38}, 356 (1983)).

\bibitem{ref4} Fern\'andes, D. J., Hussin, V. and Nieto, L. M.,
J. Phys. A: Math. Gen. {\bf 27}, 3547 (1994).

\bibitem{ref5} Kanjay, M. S. and Khare, A., Phys. Lett. A
{\bf 217}, 73 (1996).

\bibitem{ref6} Fukui, T. and Aizawa, N., Phys. Lett. A {\bf 180},
308 (1993).

\bibitem{ref7} Balantekin, A. B., Phys. Rev. A {\bf 57}, 4188 (1998).

\bibitem{ref8} Aleixo, A. N. F., Balantekin, A. B. and C\^andido 
Ribeiro, M. A., J. Phys. A: Math. Gen. {\bf 33}, 1503 (2000).

\bibitem{ref9} Aleixo, A. N. F., Balantekin, A. B. and C\^andido 
Ribeiro, M. A., J. Phys. A: Math. Gen. {\bf 33}, 3173 (2000).
  
\bibitem{ref10} Nieto, M. M. and Simmons Jr, L. M., Phys. Rev. D
{\bf 20}, 1321 (1979).

\bibitem{ref11} Nieto, M. M. and Simmons Jr, L. M., Phys. Rev. A
{\bf 19}, 438 (1979).

\bibitem{ref12} Cooper, I. L., J. Phys. A: Math. Gen. {\bf 26},
1601 (1993).

\bibitem{ref12'} Dutt, R., Khare, A. and Sukhatme, U. P., Am. J. 
Phys. {\bf 56}, 163 (1988).

\bibitem{ref14} Cooper, F., Ginocchio, J. N. and Khare, A.,
Phys. Rev. D {\bf 36}, 2458 (1987).

\bibitem{ref15} Khare, A. and Sukhatme, U. P., J. Phys. A:
Math. Gen. {\bf 26}, L901 (1993); Barclay, D., {\it et al},
Phys. Rev. A {\bf 48}, 2786 (1993).

\bibitem{ref17} Drigo Filho, E., J. Phys. A: Math. Gen. {\bf
21}, L1025 (1988).

\bibitem{ref18} Davidov, A. S., ``Quantum Mechanics'' (Pergamon, New
York, 1965), p. 131.

\bibitem{ref19} Alves, N. A. and Drigo Filho, E., J. Phys. A:
Math. Gen. {\bf 21}, 3215 (1988).

\bibitem{ref20} Nouri, S., Phys. Rev. A {\bf 60}, 1702 (1999).


\end{references}
\end{document}